# Crystal-Templating With Mutually Miscible Solvents: A Simple Path To Hierarchical Porosity

Christian Guizard[†], Jérôme Leloup, Sylvain Deville*

Laboratoire de Synthèse et Fonctionnalisation des Céramiques, UMR3080 CNRS/Saint-Gobain, Cavaillon, France

[†] now with Institut Européen des Membranes, Université de Montpellier 2, Place Eugène Bataillon, 34095 Montpellier cedex 5, France

* To whom correspondence should be addressed. E-mail: sylvain.deville@saint-gobain.com

**Abstract:** Ice templating, a route where ice crystals are used to template macroporosity, has been used to process a variety of materials with one level of macropores. We demonstrate here a variant of ice templating based on the solidification of mutually miscible solvents. The solidification of different phases, each defined by their own size and morphology, provides a simple one-step processing route for materials with hierarchical porosity defined by up to three levels of macroporosity. These concepts are demonstrated with both ceramic (yttria-stabilized zirconia) and polymer (polyvinyl alcohol) materials.

**Introduction**

Macroporous materials comprising pores at multiple scales are required for a number of applications involving mass transfer (filtration), heat transfer, or surface reactions (catalysis). Many approaches have been developed to obtain ceramic materials with a hierarchical porosity[1], such as extrusion[2] or ice templating[3]. They typically involve the presence of a macro- or mesopore former that is removed by a heat treatment. Such structures are usually defined by one level of macroporosity and meso- or microporosity. Ice templating is a well-established materials processing route suitable for making macroporous materials with a directional porosity[4]. When a solvent other than water is used, the more generic term "crystal-templating" might be used. The formation of the porosity is based on the templating effects of the solvent crystals, which thus define macroporosity on a size scale defined by the crystal size. The addition of organic pore formers[5, 6] can add an additional level of macroporosity, but a careful heat treatment step is still required to eliminate the large quantity of pore formers.

Here we demonstrate a variant of crystal-templating based on the solidification of mutually miscible solvents that is able to create macropores at several scales without the assistance of organic pore formers. The solidification of different phases, each of them defined by their own size and morphology, provide a simple one-step processing route for materials with hier-archical porosity. We demonstrate the approach in the tert-butyl-alcohol (TBA)/water system using either a ceramic (8mol.% yttria-stabilized zirconia) or a polymer (polyvinyl alcohol). TBA is a solvent extensively investigated and used in the pharmaceuticals industry[7] and its phase diagram with water (figure 1) is therefore well-known. In addition, its phase equilibria as a function of temperature and pressure are compatible with the usual ice templating conditions and equipment. The use of TBA in ice templating has already been largely documented[8–10], although always in its pure form.

**Materials and Methods**

The following products were used: distilled water, TBA (Sigma-Aldrich), 8mol.% yttria-stabilized zirconia (TZ8Y, Tosoh, Japan), dispersants (Prox B03, a polyacrylate ammonium salt from Synthron, Levallois-Paris, France), and a binder (polyethylene glycol PEG6M sold by Merck or polyvinyl-butyral (PVB) from Sigma-Aldrich) that is required to hold particles together during the freeze-drying step and avoid collapse of the porous structure. The macroporous PVA samples were obtained using a PVA (PVA LL6036) from Wacker (Burghausen, Germany). The ceramic suspensions are prepared by mixing the ceramic powder, the dispersant, the binder and the mix of solvents in predefined ratios. For TBA content up to 40wt%, ProxB03 was used as a dispersant and PEG6M as a binder. For TBA content above 70wt%, no dispersant was used and PVB was used as a binder, since it easily dissolves in concentrated TBA suspensions. The suspension was then ball-milled for 10 hours. The PVA solution was obtained by dissolving the PVA in water under magnetic stirring for 24 hrs. The suspension or solution were poured into a PTFE mold and cooled from the bottom, using a liquid-nitrogen cooled copper rod. The cooling rates were adjusted through a thermocouple and a heater. Details of the experimental setup can be found in previous papers[11]. This setup provides a fine control of the ice growth velocity, along the temperature gradient direction. Once freezing is completed, the samples were freeze-dried for at least 48 hrs. in a commercial freeze-dryer (FREE Zone 2.5Plus, Labconco, Kansas City, Missouri, USA) to ensure a complete removal of the ice and TBA (or water/TBA hydrates) crystals. Ceramic samples were densified by a high temperature sintering treatment, with a temperature rise at a rate of 180°C/h up to 500°C, holding a 500°C for 1 hr, up to 1350°C at 300°C/hr, steady stage of 3 hours at 1350°C, temperature decrease at a rate of 350°C/h to room temperature. Samples were cut, perpendicular to the direction of the temperature gradient, using a low-speed diamond saw, and cleaned in an ultrasound bath to remove the debris. SEM observations were performed using either a TM1000 from Hitachi. Pore diameters and volume were measured by mercury intrusion porosimetry (AutoPore IV 9500, Micromeritics) with applied pressures up to 200 MPa.

**Results and discussion**

Structures with a hierarchical porosity are obtained when the solidification path goes through several solidification events at different temperatures. The resulting porous structures are then defined at multiple scales (figure 2). The temperatures of the solidification events are all different, crystallization of the various phases will thus occur at different moments of the process, ensuring a progressive organization towards the final structure. The basic requirement for the solvents is that their solubility limit for particles is very low (which is the case for ice[12]), so that pure phases are grown and effectively reject all the particles in the inter-crystal space. The pores characteristics were assessed by mercury porosimetry (figure 2C). Three distinct peaks, related to the three types of pores, are clearly observed in the pore size distribution. The solidification being directional, the tortuosity of the macropores is minimal. The resulting structures have thus a hierarchical porosity with three types of macropores of different morphologies and dimensions. Finer macropores (0.5 µm) are formed by the arrangement of concentrated particles at the end of freezing. Since breakthrough of the interface in the interparticle space occurs before maximum particle packing[13], we can expect the resulting pores to be larger than expected from the maximum packing.

The porosimetry results indicate good connectivity between the various pore size scales.

The structures obtained can be rationalized by considering the phase diagram (Figure 1) and the expected sequence of solidification events, and assuming that we are constantly close to the equilibrium during solidification. If the conditions are suitable (path b in Figure 1), ice crystals form first (large pores observed in Fig. 2A and B). As solidification proceeds, the TBA content in the liquid solution increases, until it reaches the eutectic point. A eutectic structure then forms, in the intercrystal space (smaller macropores observed in Fig. 2 A and B). The expected proportion of the relative crystals can be estimated using the lever rule. The lamellar spacing in eutectic structures is typically 1/10$^{th}$ of that of dendrites grown under the same conditions[14], which seems consistent with the current observations. An equivalent behavior can be obtained in the TBA-rich side of the phase diagram, for instance with a 5wt% water/95wt% TBA zirconia suspension. A hierarchical structure with two populations of macropores is obtained (Figure 2D). The largest macropores present a facetted morphology typical of TBA-templated materials[15]. The eutectic A in the water/TBA system (trajectory c in Figure 1) yields lamellar pores with a short spacing (1-3 µm, figure 3A), a range that has been difficult to achieve thus far in the other ice templating systems[16]. The pore size and morphology are thus consistent with that observed in Figure 2. If the TBA concentration is too low (4wt% of TBA, trajectory a in Figure 1), the fraction of particle in the liquid phase rejected in the intercrystal space is too high for any templating effect to occur. The concentration of particles is so high that they cannot rearrange when the eutectic is forming. The resulting structure comprises thus just two populations of macropores (Figure 3B), created respectively by the large ice crystals and the packing of particles. The final materials comprise polycrystalline platelets of large dimensions (10–20 µm in thickness, and hundreds of microns in length and width).

A variety of polymers have successfully been ice templated previously, including PVA[17], cellulose[18], hydrogels[19], or composites[20]. The concepts demonstrated here are generic, and we successfully prepared a macroporous polyvinyl alcohol (PVA) material exhibiting two levels of macroporosity (Figure 4). Large macropores (100-200 μm wide, hundreds of microns long) are separated by macroporous walls defined comprising smaller macropores (5-30 μm wide, 20-100 μm long). Both populations of macropores are continuous along the freezing direction.

**Conclusions**

Crystal-templating based on mutually miscible solvents is thus a versatile approach to obtain porous ceramic or polymer materials with a variety of pore morphologies, including hierarchical porosity with up to three levels of macroporosity. The principles demonstrated here are generic and can easily be extended to other systems with appropriate phase diagrams. The temperature and phase separation kinetics must nevertheless be compatible with the usual ice templating temperature and pressure conditions.

**Figure**

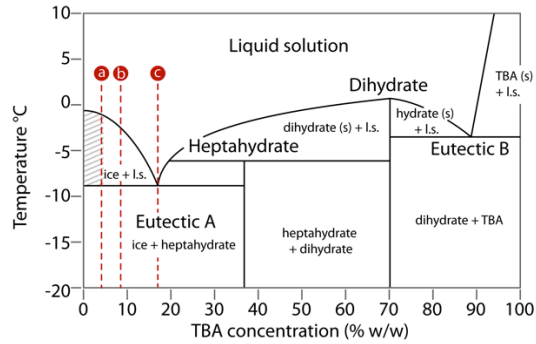

Figure 1. Phase diagram of the TBA/water system, after Vessot and Andrieu[7]. Depending on the relative concentration of water and TBA in the slurry, different types of structure may be processed, including hierarchical structures obtained with two types of crystals. The dashed area indicates the region where the TBA concentration is too low to obtain hierarchical architectures: the fraction of particle in the liquid phase rejected in the intercrystal space is too high for any templating effect to occur. The same behavior happens on the TBA-rich side of the phase diagram (dashed area not shown for clarity). The different trajectories (a, b, c) are discussed in the text.

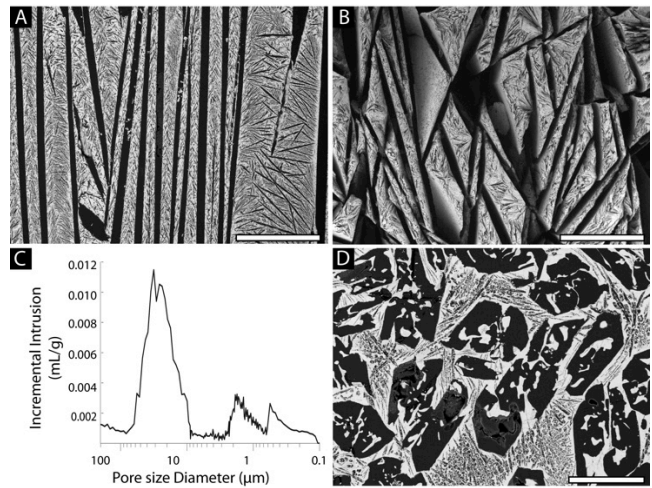

Figure 2. TZ8Y porous structure with hierarchical porosity obtained from either 84wt% water/16wt% TBA (A-B) or 5wt% water/95wt% TBA (D). SEM micrographs parallel (A) or perpendicular (B, D) to the freezing direction, and pore size distribution (C) from mercury porosimetry intrusion for the 84wt% water/16wt% TBA sample. Cooling rates: 2°C/min. The cross-sections shown in A and D were obtained after infiltration with an epoxy resin and polishing. Scale bars: A, B: 500µm, C: 150µm.

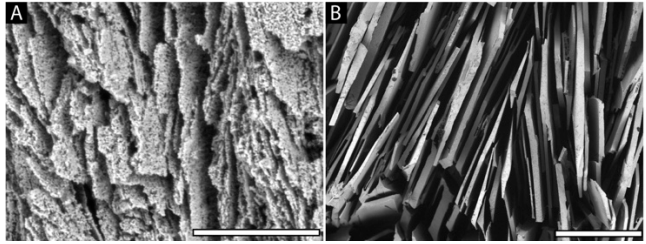

Figure 3. SEM micrographs of the different types of porous, non-hierarchical TZ8Y structures obtained. Cross-section perpendicular to the freezing direction. (A) 80wt% water/20wt% TBA (close to eutectic composition) (B) 95wt% water/5wt% TBA . Scale bars: A: 30µm, B: 500µm.

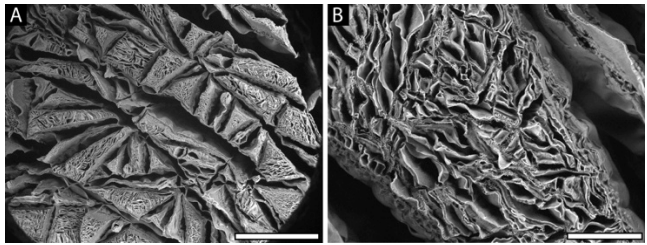

Figure 4. PVA porous structure with two levels of porosity (84wt% water/16wt% TBA). SEM micrographs perpendicular to the freezing direction. (B) Close-up view showing the finer macropores. Cooling rate: 2°C/min. Scale bars: A: 1mm, B: 150µm.